\documentclass[showpacs,preprintnumbers,amsmath,amssymb]{revtex4}
\usepackage{color}
\usepackage{comment}
\usepackage{amsmath}
\usepackage{amssymb}
\usepackage{amsthm}
\usepackage{amscd}
\usepackage{cases}
\usepackage{graphicx}
\usepackage{indentfirst}
\usepackage{epsfig,subfigure}

\newcommand\beq{\begin{equation}}
\newcommand\eeq{\end{equation}}
\newcommand\beqn{\begin{eqnarray}}
\newcommand\eeqn{\end{eqnarray}}

\newcommand\fc{\frac}
\newcommand\lt{\left}
\newcommand\rt{\right}
\newcommand\pt{\partial}

\begin{document}

\title{New localization mechanism of fermions on braneworlds }

\author{Yu-Xiao Liu$^{1,2}$\footnote{liuyx@lzu.edu.cn,
corresponding author,
}}
\author{Zeng-Guang Xu$^1$\footnote{xuzg12@lzu.edu.cn}}
\author{Feng-Wei Chen$^1$\footnote{chenfw10@lzu.edu.cn}}
\author{Shao-Wen Wei$^1$\footnote{weishw@lzu.edu.cn}}
\affiliation{Institute of Theoretical Physics,
              Lanzhou University, Lanzhou 730000, China}

\begin{abstract}
It is known that by introducing the Yukawa coupling between the fermion and the background scalar field, a bulk spin-half fermion can be localized on general Randall-Sundrum braneworlds generated by a kinklike background scalar. However, this localization mechanism does not work anymore for Randall-Sundrum braneworlds generated by a scalar whose configuration is an even function of the extra dimension. In this paper, we present a new localization mechanism for spin-half fermions for such a class of braneworld models, in which extra dimension has the topology $S^1/Z_2$. By two examples, it is shown that the new localization mechanism produces interesting results. In the first model with the brane generated by two scalars, the zero mode of the left-handed fermion is localized on the brane and there is a mass gap between the fermion zero mode and excited KK modes. In the second model with the brane generated by a dilaton scalar, the zero mode of the left- or right-chiral fermion can be localized on the brane and there is no mass gap.
\end{abstract}


\pacs{11.10.Kk., 04.50.-h.}




\maketitle

\section{Introduction}

Brane world scenarios \cite{Antoniadis,ADD,AADD,RS1,RS2},
which were motivated from string/M theory,
have been attracting continuing interest in recent years,
because they not only show a new viewpoint of spacetime,
but also provide new approaches to address a large number of outstanding issues
such as the hierarchy problem, the cosmological problem,
the nature of dark matter and dark energy,
black hole production at future colliders as a window on quantum gravity,
producing electroweak symmetry breaking without a Higgs boson and so on.
In these scenarios, our (3+1)-dimensional space-time is a submanifold (the brane)
embedded in a fundamental higher dimensional space-time (the bulk).

An important issue in braneworld theories is the mechanism by which extra dimensions
are hidden and ordinary matters are confined on the brane, so that the space-time is effectively four-dimensional at least at low energy. This can be ensured for those brane models with compact extra dimensions \cite{Antoniadis,ADD,AADD,RS1}. But in other models, extra dimensions can be infinite \cite{RS2,GRS}. For these brane models, it is interesting and important to give the mechanism of confinement of ordinary matters on the brane.


With simple field-theoretic models, in which branes are generated naturally by background scalar fields with some potential, one can investigate the localization of ordinary matters on branes. In this paper, we focus on fermions. In braneworld models, the extra dimension is usually supposed to possess $Z_2$ symmetry, hence the background scalar fields would have odd or even parity. If the scalar $\phi$ is an odd function of extra dimension, the well known localization mechanism for a fermion is to introduce the Yukawa coupling between the fermion and the background scalar field, i.e. $\eta\bar{\Psi}\phi\Psi$. There are a lot of works on this scenario ( see Refs. \cite{Volkas0705.1584,Liu0804,Lin:1992qb,Bazeia0809,Flachi0903,Fermions2010,zhaoliu2010,Kodama0812} and references therein). However, if the scalar is an even function of extra dimension, this mechanism does not work anymore, and we need to introduce new localization mechanism. This is the goal of this paper.



\section{Localization mechanism} \label{Section2}

In order to investigate the localization of spin half
fermions on the branes generated by even and/or odd background scalar fields,
we introduce a new kink of coupling between fermions and background scalars.
The five-dimensional action for a massless Dirac fermion coupled to $n$ real
scalar fields reads
\begin{eqnarray}
 S_{1/2} &=& \int d^{5}x\sqrt{-g}
    \bigg[ \bar{\Psi}\Gamma^{M}(\partial_{M}+\omega_{M})\Psi     \nonumber \\
    &&
          +\eta\bar{\Psi}\Gamma^M\partial_M F(\phi,\chi,\cdots,\rho)\gamma^5\Psi\bigg],
    \label{fermion field action}
\end{eqnarray}
where $F(\phi,\chi,\cdots,\rho)$ is a function of the real scalar fields $\phi,\chi,\cdots,$ and $\rho$, $\eta$ is the Yukawa coupling, and the spin connection $\omega_{M}$ is defined as
\begin{eqnarray}
\omega_{M}
 = \frac{1}{4}\omega^{\bar{M}\bar{N}}_{M}
         \Gamma_{\bar{M}}\Gamma_{\bar{N}}
\label{covariant derivative}
\end{eqnarray}
with
\begin{eqnarray}
\omega^{\bar{M}\bar{N}}_{M}
 &=&\frac{1}{2}E^{N\bar{M}}(\partial_{M}E^{\bar{N}}_{N}
         -\partial_{N}E^{\bar{N}}_{M})  \nonumber\\
 &-& \frac{1}{2}E^{N\bar{N}}(\partial_{M}E^{\bar{M}}_{N}
         -\partial_{N}E^{\bar{M}}_{M})\nonumber\\
 &-& \frac{1}{2}E^{P\bar{M}}E^{Q\bar{N}} E^{\bar{R}}_{M}
   (\partial_{P}E_{Q\bar{R}}-\partial_{Q}E_{P\bar{R}})\,.
   \label{SpinConnection}
\end{eqnarray}
In five dimensions, Dirac fermions are four-component spinors and their
gamma structure is determined by the gamma matrices in curved spacetime: $\{\Gamma^{M},\Gamma^{N}\}=2g^{MN}$, where
$\Gamma^{M}=E^{M}_{\bar{M}}\Gamma^{\bar{M}}$ with the
$E^{M}_{\bar{M}}$ being the vielbein and $\gamma^{\bar{M}}$ the gamma matrices in flat spacetime.
 The indices of
five-dimensional spacetime coordinates and the local lorentz
indices are labelled with capital Latin letters $M,N,\ldots$ and
$\bar{M},\bar{N},\ldots$, respectively.

The metric describing a Minkowski brane in five dimensions is assumed as \cite{RS1}
\begin{equation}
ds^{2}_{5}=e^{2A(y)}\eta_{\mu\nu}dx^{\mu}dx^{\nu}+dy^{2} ,
\label{FlatBraneMetric}
\end{equation}
which can also be transformed to the conformally flat one
\begin{equation}
ds^{2}_{5}=e^{2A(z)}(\eta_{\mu\nu}dx^{\mu}dx^{\nu}+dz^{2})
\label{conformallyFlatMetric}
\end{equation}
by the coordinate transformation
\begin{equation}
dz=e^{-A(y)}dy, \label{coordinateTransformation}
\end{equation}
where $\text{e}^{2A}$ is the warp factor, $y$ or $z$ denotes the extra dimension coordinate, and $\eta_{\mu\nu}$ is the induced metric on the brane.
With the conformally flat metric (\ref{conformallyFlatMetric}),
we will conveniently get the corresponding Schr\"{o}dinger-like equations and
mass-independent potentials for left-chiral and right-chiral fermion KK modes.

With the conformally flat metric (\ref{conformallyFlatMetric}), the spin connection $\omega_{M}$ read as
$\omega_{\mu}=\frac{1}{2}({\partial_z}A(z))\gamma_{\mu}\gamma_{5}$ and $\omega_5=0$, and the equation of motion for the five-dimensional Dirac fermion is
\begin{eqnarray}
\Big[\gamma^{\mu}\partial_{\mu}
     +\gamma^5 \big(\partial_z+2{\partial_z}A(z)\big)
     +\eta \partial_z F \Big]\Psi=0.\label{DiracEq1}
\end{eqnarray}
Note that for the usual Yukawa coupling $-{\eta}F(\phi,\chi,\cdots,\rho)\bar{\Psi}\Psi$, the corresponding Dirac equation is given by \cite{Volkas0705.1584,Liu0804,Bazeia0809,Flachi0903,Fermions2010,zhaoliu2010,Kodama0812,two-scalarbrane1,Fu:2011pu,Correa:2010zg}
\begin{eqnarray}
\Big[\gamma^{\mu}\partial_{\mu}
    +\gamma^5 \big(\partial_z+2{\partial_z}A(z)\big)
    +\eta e^{A(z)} F \Big]\Psi=0.\label{DiracEq1Yukawa}
\end{eqnarray}

In order to investigate the above Dirac equation (\ref{DiracEq1}), we make the general chiral decomposition in terms of four-dimensional effective Dirac fields:
\begin{eqnarray}
 \Psi(x,z) &=&
       \sum_{n}\psi_{Ln}(x)f_{Ln}(z) e^{-2A(z)} \nonumber \\
   &+& \sum_{n}\psi_{Rn}(x)f_{Rn}(z) e^{-2A(z)}
   ,
\label{the general chiral decomposition}
\end{eqnarray}
where $\psi_{Ln}(x)=-\gamma^{5}\psi_{Ln}(x)$ and
$\psi_{Rn}(x)=\gamma^{5}\psi_{Rn}(x)$ are the left- and
right-chiral components of the four-dimensional effective Dirac fermion
field, respectively, and they satisfy the four-dimensional massive
Dirac equations
$\gamma^{\mu}\partial_{\mu}\psi_{Ln}(x)=\mu_{n}\psi_{Rn}(x)$ and
$\gamma^{\mu}\partial_{\mu}\psi_{Rn}(x)=\mu_{n}\psi_{Ln}(x)$. Note that the summation indices are not necessarily the same for the left- and right-chiral KK modes. In the following, we mainly focus on the left- and
right-chiral KK modes $f_{Ln}(z)$ and $f_{Rn}(z)$ of the five-dimensional Dirac field, which satisfy the following coupled equations:
\begin{eqnarray}
&&[\partial_z-\eta\partial_z F]f_{Ln}(z)=+\mu_n f_{Rn}(z), \label{spinorCoupledEqs1}\\
&&[\partial_z+\eta\partial_z F]f_{Rn}(z)=-\mu_n f_{Ln}(z), \label{spinorCoupledEqs2}
\end{eqnarray}
which can be recast into
\begin{subequations}\label{SuperSymmetricForm}
\begin{eqnarray}
&&U^{\dagger}U f_{Ln}(z)=\mu_n^2 f_{Ln}(z), \\
&&U U^{\dagger}f_{Rn}(z)=\mu_n^2 f_{Rn}(z),
\end{eqnarray}
\end{subequations}
with the operator $U$ defined as $U=\partial_z-\eta\partial_z F(\phi)$.
The above equations can also be rewritten as the Schr\"{o}dinger-like equations
\begin{subequations}\label{Scheq}
\begin{eqnarray}
 [-\partial_{z}^{2}+V_{L}(z)]f_{L}(z) &=& \mu_{n}^{2}f_{L}(z) \,,
    \\ \label{ScheqLeft}
 [-\partial_{z}^{2}+V_{R}(z)]f_{R}(z) &=& \mu_{n}^{2}f_{R}(z) \,,
       \label{ScheqRight}
\end{eqnarray}
\end{subequations}
where the effective potentials for the KK
modes $f_{L,R}$ are
\begin{eqnarray}
V_{L,R}(z)=\big(\eta\partial_z F\big)^2
        \pm \partial_{z}\big(\eta\partial_z F\big).  \label{VzLR}
\end{eqnarray}
The form of Eq.~(\ref{SuperSymmetricForm}) and the supersymmetric partner potentials (\ref{VzLR}) show that there is no tachyon fermion KK modes with negative mass square, and it also indicates that we may obtain a chiral massless fermion on the brane for the new coupling.
For the usual Yukawa coupling $-{\eta}F(\phi,\chi,\cdots,\rho)\bar{\Psi}\Psi$, the corresponding effective potentials read \cite{Volkas0705.1584,Liu0804,Bazeia0809,Flachi0903,Fermions2010,zhaoliu2010,Kodama0812,two-scalarbrane1,Fu:2011pu,Correa:2010zg}
\begin{eqnarray}
V_{L,R}(z)=\big(\eta e^A  F\big)^2
        \pm \partial_{z}\big(\eta e^A F\big).  \label{VzLROddScalar}
\end{eqnarray}

On the other hand, in order to derive the effective action on the brane for the four-dimensional
massless and massive Dirac fermions, we need the following orthonormality
conditions for the KK modes $f_{Ln}(z)$ and $f_{Rn}(z)$:
\begin{eqnarray}
 \int \!\! f_{Lm}f_{Ln}dz=
 \int \!\! f_{Rm}f_{Rn}dz=\delta_{mn}\,
 \int \!\! f_{Lm}f_{Rn}dz=0,
  \label{OrthonormalityConditions}
\end{eqnarray}
which are the same as case of the usual Yukawa coupling.

The zero modes of the left- and right-chiral fermions are turned out to be
\begin{eqnarray}
 f_{L0,R0}(z) \propto
  \exp\bigg[\pm
   \int_{0}^{z} d\bar{z}
        \eta \partial_{\bar{z}} F
         \bigg]
  = \exp\big(\pm\eta  F \big).     \label{fL0CaseI}
\end{eqnarray}
The normalization condition is
\begin{eqnarray}
 \int \exp\big(\pm2\eta  F \big) dz < \infty.
       \label{fL0CaseI}
\end{eqnarray}
So, there is at most one of the left- and right-chiral fermion zero modes that can be localized on the brane if the extra dimension in the conformal coordinate $z$ is infinite. We can obtain the solutions of right-chiral KK modes from the left-chiral ones with the replacement $\eta \rightarrow-\eta$. So we mainly focus on $\eta>0$.

In general, the extra dimension has $Z_2$ symmetry for a brane model, so the effective potentials for the fermion KK modes should be symmetric with respect to extra dimension.
In order to ensure the potentials $V_L(z)$ and $V_R(z)$ are even functions of $z$, $F$ must be an even function of $z$ according to Eq. (\ref{VzLR}). So, if the background scalars are even, the usual Yukawa coupling $-{\eta}F(\phi,\chi,\cdots,\rho)\bar{\Psi}\Psi$ does not work for the fermion localization, and we need the new one introduced here.

Next, we will investigate localization of fermions on various branes by using the new localization mechanism presented here. We will give two examples.



\section{Localizations of fermions on branes}

\subsection{Two-field thick brane}

We first investigate the localization of fermions on a two-field thick brane. The system is described by the action including two interacting scalars $\phi$ and $\pi$:
\begin{equation}\label{action}
 S=\int d^{5}x\sqrt{-g}\bigg[\frac{1}{2\kappa_5^2}R-\frac{1}{2}(\partial\phi)^{2}
   -\frac{1}{2}(\partial\pi)^{2}-V(\phi,\pi)\bigg],
\end{equation}
where $R$ is the scalar curvature. We set $\kappa_5^2=8 \pi G_5=1$ with $G_5$ the 5-dimensional Newton constant.
The line element is also assumed as (\ref{conformallyFlatMetric}) for a static Minkowski brane.
The background scalars $\phi$ and $\pi$ are only dependent on the extra coordinate $z$. Then the field equations read
\begin{eqnarray}
 \phi'^2 + \pi'^2  &=& 12A'^2 +2\text{e}^{2A} V, \label{EOM1} \\
  \phi'^2 + \pi'^2  &=& 3A'^2-3A'', \label{EOM2} \\
 \phi'' + 3A'\phi' &=& \text{e}^{2A}\frac{\partial V}{\partial \phi}, \label{EOM3} \\
 \pi'' + 3A'\pi' &=& \text{e}^{2A}\frac{\partial V}{\partial \pi},\label{EOM4}
\end{eqnarray}
where the prime denotes the derivative with respect to $z$. Note that only three of the above equations are independent because of the conservation of the energy-momentum tensor.

The equations (\ref{EOM1})-(\ref{EOM4}) can be solved via the superpotential method \cite{DeWolfe}. Here we set $V= \frac{1}{2}\text{e}^{-2\sqrt{1/3}\;\pi}\left[(\frac{d W(\phi)}{d\phi})^{2}-W(\phi)^{2}\right]$, then the equations are transformed as the following ones:
\begin{eqnarray}
\phi'=\frac{d W}{d \phi}, ~~~A' = -\frac{1}{3}W,~~ \pi = \sqrt{3} A.\label{solution1}
\end{eqnarray}
For a specific superpotential $W(\phi)$
\cite{two-scalarbrane1,KRthickbrane}:
\begin{equation}
W(\phi)=va\phi\left(1-\frac{\phi^{2}}{3v^{2}}\right),
\end{equation}
the solution is \cite{Liu:2011ysa,Fu:2011pu}
\begin{eqnarray}
\phi(z)&=&v\tanh(az),\label{dilat}\\
A(z)&=&-\frac{v^2}{9}\Big[\ln\cosh^2(az)
                        +\frac{1}{2}\tanh^2(az)\Big], \label{warpfactor}\\
\pi(z)&=&\sqrt{3} \;A(z),\label{dilat2}
\end{eqnarray}
where both $v$ and $a$ are positive constants. It can be seen that the solution for $\phi$ is a kink and $\pi=\sqrt{3b}A$ is a dilaton field.

In order to localize fermions on the brane, we need consider the coupling between fermions and background scalars. If we consider the usual Yukawa coupling between fermions and the kink, i.e., $-\eta\bar{\Psi}\phi\Psi$, then we will find that fermion zero modes can not localized on the brane \cite{Fu:2011pu}. If we choose the coupling $-\eta\bar{\Psi}\text{e}^{\lambda\pi}\phi\Psi$, then one of the left- and right-chrial fermion zero modes, namely the four-dimensional left- or right-chrial massless fermion, is localized on the brane for $\lambda\leq-1/\sqrt{3}$. And there will be finite and infinite number of four-dimensional massive fermions localized on the brane for $\lambda=-1/\sqrt{3}$ and $\lambda<-1/\sqrt{3}$, respectively \cite{Fu:2011pu}.

\begin{figure}[htb]
\begin{center}
\subfigure[$\eta=1.6$]{
\includegraphics[width=5cm,height=3cm]{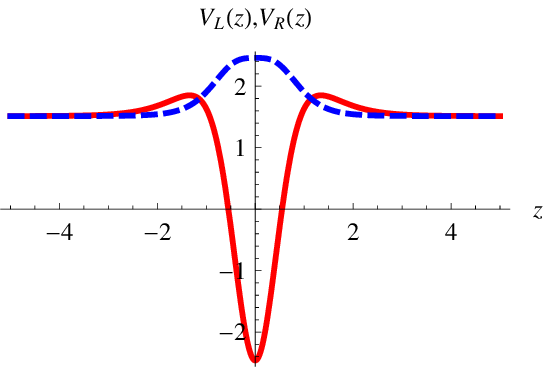}}
\subfigure[$\eta=2.6$]{
\includegraphics[width=5cm,height=3cm]{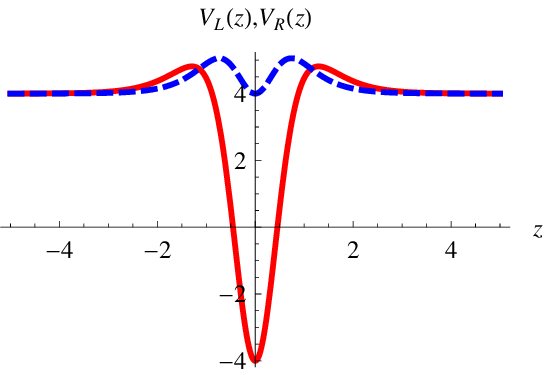}}
\subfigure[$\eta=6.6$]{
\includegraphics[width=5cm,height=3cm]{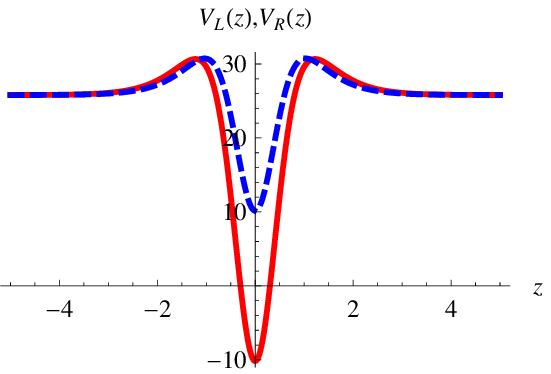}}
\end{center}
\caption{The shapes of the potentials $V_{L}(z)$ (red curves) and $V_{R}(z)$ (blue dashed curves) for the two-field brane.
The parameters are set to $a=1$, $v=2$, and $\eta=1.6,~2.6,~6.6$.
}
\label{VLVR1}
\end{figure}

Here, we ask an interesting question: can fermions be localized on the brane if they only couple with the even background scalar, i.e., the dilaton field $\pi$? In order to answer this question, we apply the localization mechanism developed in previous section and consider the simplest case $F=\pi$, i.e., with the coupling of $\eta\bar{\Psi}\Gamma^M(\partial_M \pi) \gamma^5\Psi$, for which the potentials (\ref{VzLR}) are given by
\begin{eqnarray}
V_{L,R}(z)
        &=&\frac{1}{54} a^2 v^2 \eta
           \Big[2v^2 \eta
                \left(1+{\text{sech}^{2}(az)}\right)^2 \tanh^{2}(az)  \nonumber \\
        &&\pm 3\sqrt{3}\big(\cosh(2az)-5\big){\text{sech}^{4}(az)}\Big]. \label{VzLR1}
\end{eqnarray}
The values of $V_L(z)$ and $V_R(z)$ at $z=0$ and $z\rightarrow\pm\infty$ are given by
\begin{eqnarray}
V_{L,R}(0)&=& \mp\frac{2}{3\sqrt{3}} a^2 v^2 \eta,   \nonumber \\
V_{L,R}(\pm\infty) &=& \frac{1}{27} a^2 v^4 \eta^2. \label{VzLR1}
\end{eqnarray}
Here, we only consider positive coupling constant $\eta$, for which we have $V_{L}(0)<0$ and $V_{R}(0)>0$.
Since the value of the potential $V_L(z)$ is positive at the boundary along extra dimension,
there is a mass gap, and those left-chiral fermion KK modes (including the zero mode) with $m_n^2<\frac{1}{27} a^2 v^4 \eta^2$ belong to discrete spectrum and those with $m_n^2>\frac{1}{27} a^2 v^4 \eta^2$ belong to continues one.
For right-chiral fermion KK modes, the zero mode can not be localized on the brane, and the spectrum is decided by the value of the coupling constant $\eta$. For $0<\eta<6\sqrt{3}/v^2$, $V_R(0)> V_R(\pm\infty)$, there are not bound right-chiral fermion KK modes, namely, no right-chiral fermions can be localized on the brane. If $\eta>6\sqrt{3}/v^2$, there may exist finite number of bound right-chiral fermion KK modes, whose masses must also satisfy $m_n^2<\frac{1}{27} a^2 v^4 \eta^2$.
The shapes of $V_{L}(z)$ and $V_{R}(z)$ are shown in Fig.~\ref{VLVR1} for three kinds of values of $\eta$: $0<\eta<6\sqrt{3}/v^2$, $\eta=6\sqrt{3}/v^2$, and $\eta>6\sqrt{3}/v^2$.

The left-chiral fermion zero mode reads
\begin{eqnarray}
f_{L0}(z)\propto \text{sech}^{-\frac{{\eta}v^2}{3\sqrt{3}}}(az) \exp\left( -\frac{{\eta}v^2}{6\sqrt{3}} \tanh^2(az) \right). \label{fL0CaseI}
\end{eqnarray}
It is easy to show that the zero mode (\ref{fL0CaseI}) is normalizable, so it is localized on the brane. While the right-chiral fermion zero mode is divergent (for positive coupling $\eta$) at the boundary of extra dimension and can not be localized.

The massive KK modes can be solved numerically, but we do not discuss them here.

\subsection{Scalar-tensor brane}

Next, we turn to another brane scenario---scalar-tensor brane.

In the RS1 model \cite{RS1}, there are two 3-branes located at the boundaries of a compact extra dimension with the topology $S^1/Z_2$. In order to solve the gauge hierarchy problem by the exponential warp factor $e^{-ky}$, our universe should be located on the negative tension brane. However, this would give a ``wrong-signed" Friedmann-like equation, which leads to a severe cosmological problem \cite{Csaki1999,Cline1999}. In the RS2 model, the cosmological problem has been solved, but the gauge hierarchy problem is left.

Recently, a simple generation of the RS1 model in the scalar-tensor gravity was given in Ref. \cite{Yang:2012dd}. In this model, our world is moved to the positive tension brane but the hierarchy problem is also solved.
The action for the scalar-tensor gravity is given by \cite{Yang:2012dd}
\beq
S_5=\fc{M_{5}}{2}\int{d^5x\sqrt{|g|}e^{k\phi}\lt[R-(3+4k)(\pt\phi)^2\rt]},\label{Action}
\eeq
where $M_{5}$ is the five-dimensional scale of gravity, and $k$ is a coupling constant. The braneworld is generated by the scalar $\phi$. For the special case of $k=-1$, the above action is just the standard bosonic part of the effective string action involving only the metric and the dilaton.

The line element is given by (\ref{conformallyFlatMetric}) for a static Minkowski brane.
The conformal coordinate $z\in[-z_b,z_b]$ denotes an $S^1/Z_2$ orbifold extra dimension.
This system has two sets of brane solutions. The first one is given by \cite{Yang:2012dd}
\beqn
e^{A(z)}&=&(1+\beta|z|)^{\fc{1}{3+2k}},\label{Warpfactor_2a} \\
\phi(z)&=&\fc{2}{3+2k}\ln(1+\beta|z|), \label{Scalar_2a}
\eeqn
where $\beta>0$ and $k<-{3}/{2}$.
The second solution read \cite{Yang:2012dd}
\begin{eqnarray}
e^{A(z)}&=&(1+\beta |z|)^{\fc{3+4k}{9+6k}},\label{Warpfactor_2b}\\
\phi(z)&=&-\fc{2}{3+2k}\ln(1+\beta|z|),\label{Scalar_2b}
\end{eqnarray}
where $\beta>0$ and $-3/2<k<-3/4$.

Both solutions describe the same braneworld picture: there are a positive tension brane at the origin and a negative one at the boundary $z_b$, which is similar to the RS1 model. However, the massless graviton for both solutions here is localized on the negative tension brane, which is opposite to the case of the RS1 model. Then it can be shown that, if we suppose that the Standard Model fields are confined on the positive tension brane localized at $z=0$, which is crucial to overcome the severe cosmological problem of the RS1 model, the gauge hierarchy problem can be solved in this brane model \cite{Yang:2012dd}.

Now, we would like to investigate the localization of fermions on the scalar-tensor brane. We first note that, since the scalar field $\phi(z)$ in this brane model is an even function of $z$, we can not use the Yukawa coupling with the form $-\eta\bar{\Psi}F(\phi)\Psi$, which would lead to an odd effective potential.
We would like to show that, if we apply the new localization mechanism presented in Section \ref{Section2}, fermions can be localized on the positive tension brane even the extra dimension is extended to $-\infty < z < \infty$.
Here, we consider the simple case $F(\phi)=\phi^q$ with $q$ a positive integer and $\phi$ given by the first solution (\ref{Scalar_2a}), for which the potentials (\ref{VzLR}) are given by
\begin{eqnarray}
V_{L,R}(z)&=&\frac{4q\beta^2  \phi^{q-2}(z)}{(3+2k)^2 (1+\beta |z|)^2}
             \Big[q \phi^q(z)\eta^2 \nonumber \\
           &&
             \pm \big(q-1-\ln(1+\beta |z|)\big)\eta \Big] \nonumber \\
           &&+\frac{4\beta q}{(3+2k)}\delta_{q,1}\delta(z). \label{VzLR2a}
\end{eqnarray}
For the simplest case $q=1$, the above potentials are simplified as
\begin{eqnarray}
 V_{L,R}(z)=\frac{2\beta^2 [2\eta^2\mp(2+3k)\eta]}{(3+2k)^2 (1+\beta |z|)^2}
             +\frac{4\beta q}{(3+2k)}\delta(z), ~(q=1),
\end{eqnarray}
which are just volcano potentials with a negative $\delta(z)$ potential well.
When $q\geq2$, there is no $\delta(z)$ term in the potentials anymore.

The values of $V_{L,R}(z)$ at $z\rightarrow0$ and $z\rightarrow\pm\infty$ are given by
\begin{eqnarray}
V_{L,R}(z\rightarrow0)& \rightarrow &
  \left\{\begin{array}{ll}
        +\frac{4\beta^2\eta^2}{(3+2k)^2}
            +\frac{4\beta q}{(3+2k)}\delta(z), & q=1 \\
        \pm\frac{8\beta^2\eta}{(3+2k)^2}, & q=2 \\
        0, & q\geq3
\end{array}\right.,   \nonumber \\
V_{L,R}(z\rightarrow\pm\infty) &\rightarrow& 0. \label{VzLR1}
\end{eqnarray}

\begin{figure}[htb]
\begin{center}
\subfigure[$q=1$]{
\includegraphics[width=5cm]{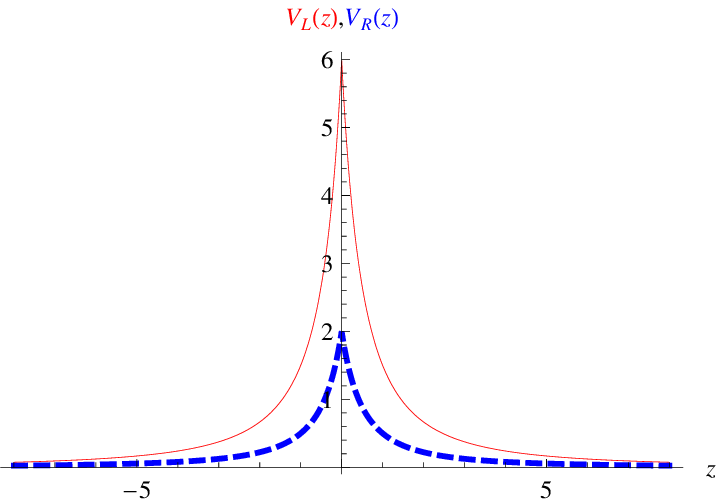}
\includegraphics[width=5cm]{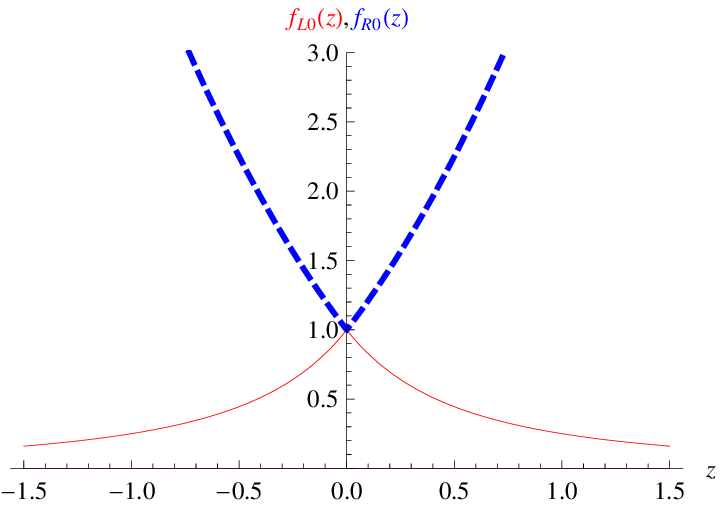}}
\subfigure[$q=2$]{
\includegraphics[width=5cm]{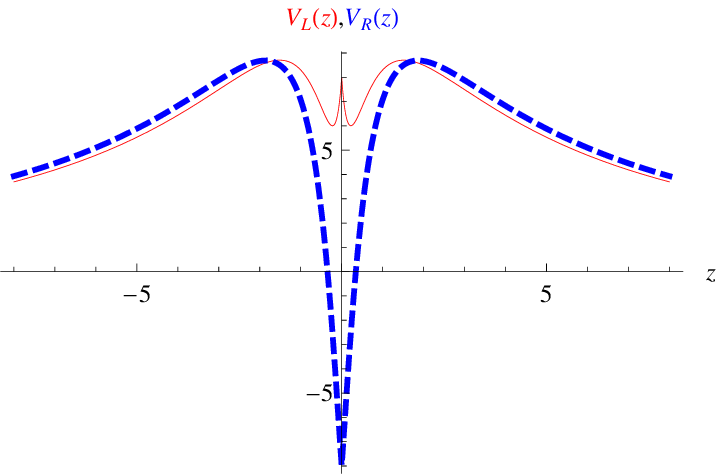}
\includegraphics[width=5cm]{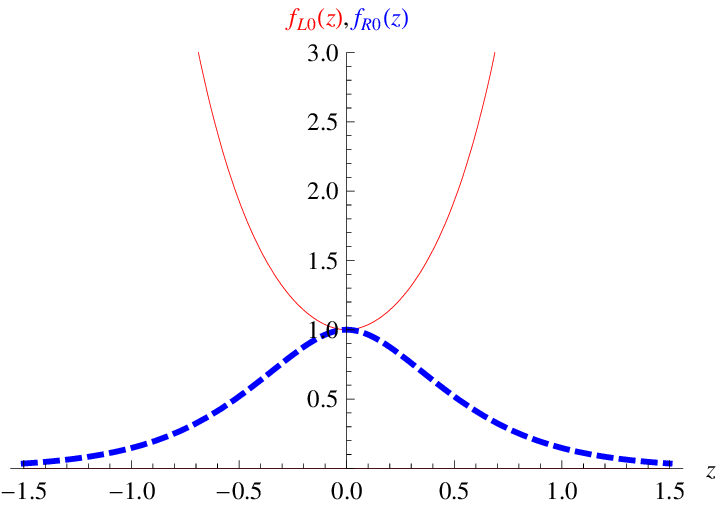}}
\subfigure[$q=3$]{
\includegraphics[width=5cm]{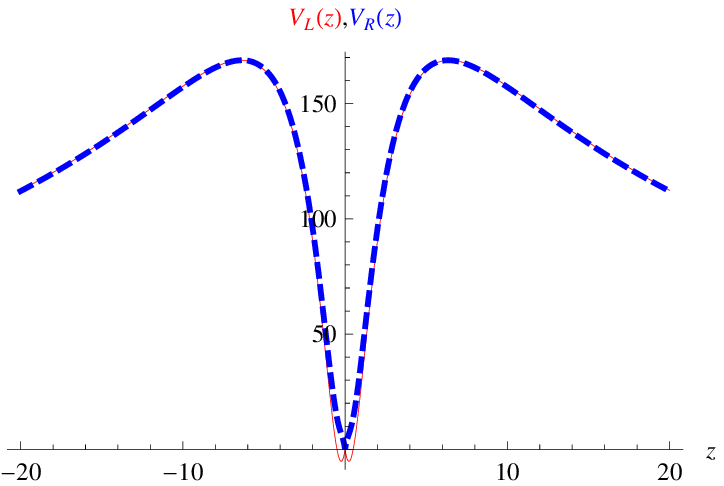}
\includegraphics[width=5cm]{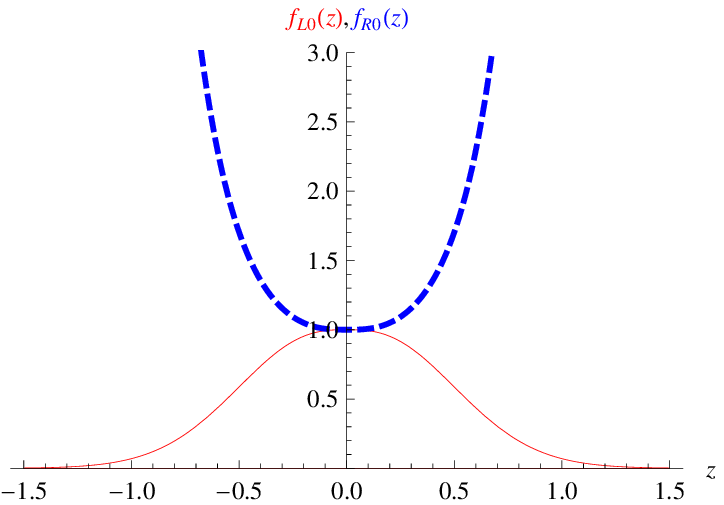}}
\subfigure[$q=4$]{
\includegraphics[width=5cm]{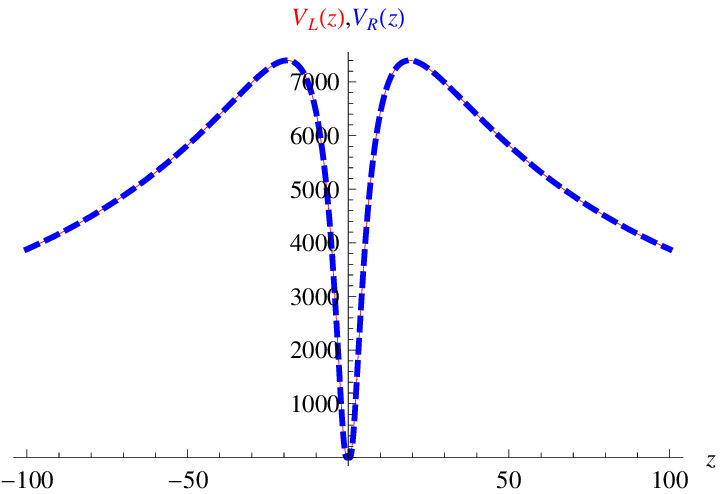}
\includegraphics[width=5cm]{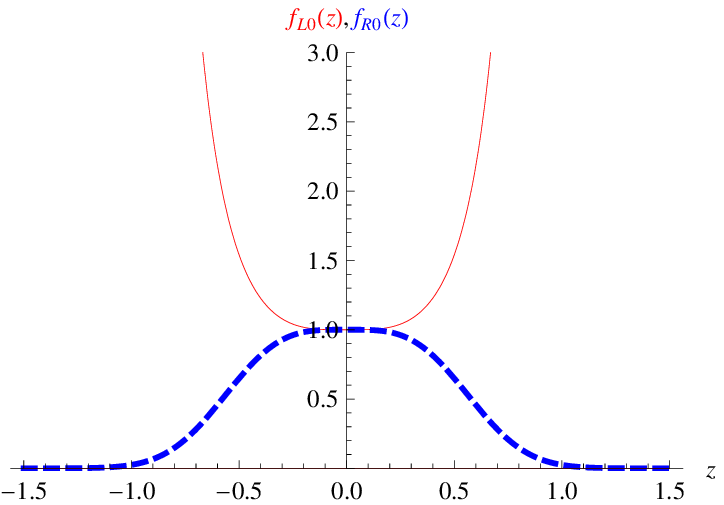}}
\end{center}
\caption{The shapes of the potentials $V_{L}(z)$ (red curves) and $V_{R}(z)$ (blue dashed curves), and the corresponding fermion zero modes $f_{L0}(z)$ (red curves) and $f_{R0}(z)$ (blue dashed curves) for the scalar-tensor brane.
The parameters are set to $\beta=1$, $k=-2$, $\eta=1$, and $q=1,2,3,4$ from top to bottom.
}
\label{Fig_VLVR2}
\end{figure}

The shapes of the potentials are plotted in Fig. \ref{Fig_VLVR2}, form which it can be seen that there are two potential barriers around the positive tension brane, and the potential barriers trend to vanish at the boundaries of extra dimension. For the case $q=1$, both potentials for left- and right-chiral fermion KK modes has a negative $\delta(z)$ potential well. For odd $q>1$, the potential $V_L(z)$ around the positive tension brane is negative, which leads to a bound left-chiral fermion zero mode. For even $q>1$, the potential $V_R(z)$ around the positive tension brane is negative, which results in a bound right-chiral fermion zero mode.

The fermion zero modes are
\begin{eqnarray}
 f_{L0,R0}(z)&\propto&   \exp\left[\pm\eta \left(\fc{2}{3+2k}\ln(1+\beta|z|)\right)^q \right].     \label{fL0CaseII}
\end{eqnarray}
The normalization conditions are given by
\begin{eqnarray}
\int_{-\infty}^{\infty}f^2_{L0,R0}(z) dz < \infty .
\end{eqnarray}
For $q=1$, if $\eta>-\frac{3+2k}{4}(>0)$, the above integral is finite, so the zero mode of left-chiral fermion can be localized on the positive tension brane. For odd $q>1$ and even $q>1$, with positive coupling $\eta$, the zero modes of the left- and right-chiral fermions are localized on the brane, respectively. The zero modes are plotted in Fig. \ref{Fig_VLVR2} for $q\leq4$.

For the solution (\ref{Scalar_2b}), the results are almost the same except  except that the localization condition of left-chiral fermion zero mode is $\eta>\frac{3+2k}{4}(>0)$ for the case $q=1$.

When $z_b<\infty$, the localizations of the left- and right-chiral fermion the zero modes are opposite, namely, one is localized on the positive tension brane and another is localized on the negative tension one.

\section{Discussion and Conclusion}

In this paper, we have presented a new localization mechanism for fermions in a class of braneworld models, in which extra dimension has the topology $S^1/Z_2$. This new localization mechanism is necessary for those braneworlds generated only by a dilaton scalar. In such braneworld models, the background scalar is an even function of extra dimension. Therefore, the usual localization mechanism, by introducing the Yukawa coupling between the fermion and background scalar, can not work anymore, because the effective potentials for fermion KK modes are not even functions of extra dimension. While the new localization mechanism introduced in the paper will give good results.

We have illustrated this by two examples. The first example is about a brane generated by two scalar fields with interaction potential, one is the usual kink scalar and another is the dilaton. For this model, our new localization mechanism gives very good and interesting result: the zero mode of left-handed fermion is localized on the brane, there is a mass gap between the fermion zero mode and excited KK modes, and there are some bound discrete fermion KK modes and a series of continue fermion KK modes.

In the second example, we consider a brane generated by a dilaton scalar. This model is a simple generation of the RS1 model in the scalar-tensor gravity. In this model, our world is moved to the positive tension brane, and the hierarchy problem and cosmological problem can be also solved synchronously \cite{Yang:2012dd}. In order to localize fermions on the positive tension brane, we considered the new coupling introduced in this paper with $F(\phi)=\phi^q$ and positive coupling constant. It was found that the zero modes of the left- and right-chiral fermions can be localized on the brane for odd and even positive integer $q$, respectively. There are a series of continue gapless massive fermion KK modes besides the zero mode.

\section*{Acknowledgement}

This work was supported by the National Natural Science Foundation of China (Grants No. 11075065, No. 11205074, and Grant No. 11375075), and the Fundamental Research Funds for the Central Universities (Grants No. lzujbky-2013-18 and Grant No. lzujbky-2013-21).


\begin{thebibliography}{10}

%

\bibitem{Antoniadis}
I. Antoniadis, {\em``A possible new dimension at a few TeV''}, Phys. Lett. \textbf{B 246} 377 (1990).

\bibitem{ADD}
N. Arkani-Hamed, S. Dimopoulos and G. Dvali, {\em``The hierarchy problem and new dimensions at a millimeter''}, Phys. Lett. \textbf{B 429} 263 (1998), [arXiv:hep- ph/9803315].

\bibitem{AADD}
I. Antoniadis, N. Arkani-Hamed, S. Dimopoulos and G.R. Dvali, {\em``New dimensions at a millimeter to a Fermi and superstrings at a TeV''}, Phys. Lett. \textbf{B 436} 257 (1998), [arXiv:hep-ph/9804398].


\bibitem{RS1}
L. Randall and R. Sundrum, {\em``A Large Mass Hierarchy from a Small Extra Dimension''}, Phys. Rev. Lett. \textbf{83} 3370 (1999), [arXiv:hep-ph/9905221].

\bibitem{RS2}
L. Randall and R. Sundrum, {\em``An alternative to compactification''}, Phys. Rev. Lett. \textbf{83} 4690 (1999), [arXiv:hep-th/9906064].

\bibitem{GRS}
R. Gregory, V. A. Rubakov and S. M. Sibiryakov, {\em``Opening up extra dimensions at ultra-large scales''}, Phys. Rev. Lett. \textbf{84} 5928 (2000), [arXiv:hep-th/0002072].


\bibitem{Volkas0705.1584}
 T. R. Slatyer and R. R. Volkas,
    J. High Energy Phys. \textbf{04}, 062 (2007).



\bibitem{Liu0804}
 Y.-X. Liu, L.-D. Zhang, L.-J. Zhang and Y.-S. Duan, {\em``Fermions on Thick Branes in the Background of Sine-Gordon Kinks''}, Phys. Rev. \textbf{D 78} 0650 25 (2008), [arXiv:0804.4553[hep-th]].

\bibitem{Lin:1992qb}
L.~Lin, G.~Munster, M.~Plagge, I.~Montvay, H.~Wittig, C.~Frick and T.~Trappenberg, {\em``Mass spectrum and bounds on the couplings in Yukawa models with mirror fermions''}, Nucl.\ Phys.\ Proc.\ Suppl.\  {\bf 30}, 647 (1993) [hep-lat/9212015].




\bibitem{Bazeia0809}
D. Bazeia, F.A. Brito and R.C. Fonseca,
    {\em``Fermion states on domain wall junctions and the flavor number''},
    Eur. Phys. J. \textbf{C 63} 163 (2009), [arXiv:0809.3048 [hep-th]].


\bibitem{Flachi0903}
A. Flachi and M. Minamitsuji,
    {\em``Field localization on a brane intersection in anti-de Sitter spacetime''},
    Phys. Rev. \textbf{D 79} 104021 (2009), [arXiv:0903.0133 [hep-th]].


\bibitem{Fermions2010}
A.~E.~R.~Chumbes, A.~E.~O.~Vasquez and M.~B.~Hott,
    ``{\em Fermion localization on a split brane}'',
    Phys.\ Rev.\ D {\bf 83}, 105010 (2011)  [arXiv:1012.1480 [hep-th]].




\bibitem{zhaoliu2010}
 Z.-H. Zhao, Y.-X. Liu, H.-T. Li and Y.-Q. Wang,
    {\em``Effects of the variation of mass on fermion localization and resonances on thick branes''},
    Phys. Rev. \textbf{D 82 } 084030 (2010), [arXiv:1004.2181[hep-th]].



\bibitem{Kodama0812}
 Y. Kodama, K. Kokubu and N. Sawado,
    {\em``Localization of massive fermions on the baby-skyrmion branes in 6 dimensions''},
    Phys. Rev. \textbf{D 79} 065024 (2009), [arXiv:0812.2638[hep-th]].

%


\bibitem{two-scalarbrane1}
A. Kehagias and K. Tamvakis,
    {\em``Localized Gravitons, Gauge Bosons and Chiral Fermions in Smooth Spaces Generated by a Bounce''},
    Phys. Lett. \textbf{B 504} 38 (2001), [arXiv:hep-th/0010112].


\bibitem{Fu:2011pu}
C.-E. Fu, Y.-X.~Liu and H.~Guo,
    {\em``Bulk matter fields on two-field thick branes''},
    Phys.\ Rev.\ D {\bf 84}, 044036 (2011), [arXiv:1101.0336 [hep-th]]

\bibitem{Correa:2010zg}
 R.~A.~C.~Correa, A.~de Souza Dutra and M.~B.~Hott,
    {\em``Fermion localization on degenerate and critical branes''},
    Class.\ Quant.\ Grav.\  {\bf 28}, 155012 (2011)  [arXiv:1011.1849 [hep-th]].


\bibitem{DeWolfe}
O. DeWolfe, D.Z. Freedman, S.S. Gubser and A. Karch,
    {\em``Modeling the fifth dimension with scalars and gravity''},
    Phys. Rev. \textbf{D 62} 046008 (2000), [arXiv:hep-th/9909134].

\bibitem{KRthickbrane}
 M.O. Tahim, W.T. Cruz and C.A.S. Almeida,
    {\em``Tensor gauge field localization in branes''},
    Phys. Rev. \textbf{D 79} 085022 (2009), [arXiv:0808.2199[hep-th]].

\bibitem{Liu:2011ysa}
Y.-X.~Liu, C.-E.~Fu, H.~Guo and H.-T.~Li,
    {\em``Deformed brane with finite extra dimension''},  Phys.\ Rev.\ D {\bf 85}, 084023 (2012), [arXiv:1102.4500 [hep-th]].

\bibitem{Csaki1999}
C.~Csaki, M.~Graesser, C.~F. Kolda, and J.~Terning,
    { Phys. Lett.} {\bf B 462} (1999) 34,
    [arXiv:hep-ph/9906513].

\bibitem{Cline1999}
J.~M. Cline, C.~Grojean, and G.~Servant,
    { Phys. Rev. Lett.} {\bf 83} (1999) 4245,
    [arXiv:hep-ph/9906523].

\bibitem{Yang:2012dd}
 K.~Yang, Y.-X.~Liu, Y.~Zhong, X.-L.~Du and S.-W.~Wei,
    {\em``Gravity localization and mass hierarchy in scalar-tensor branes''}, Phys.\ Rev.\ D {\bf 86}, 127502 (2012)  [arXiv:1212.2735 [hep-th]].



\end{thebibliography}
\end{document}